# On Detecting Multiple Simultaneous Change-points in High Dimensional Non-Stationary Time Series

**Richard Song**

University of North Florida
ssoonngg123@gmail.com

## Abstract

This paper studies the detection of multiple simultaneous (systematic) change points for high-dimensional nonstantionary economic and financial time series data. The analytic framework used is based on the standard and adaptive fused group lasso method, where the mixed $L_{2,1}$ penalty is either uniform or re-weighted by data-dependent weights. This paper shows that, under appropriate conditions, this approach is $L_2$ consistent and, by adopting the data-dependent weights, could correctly select the change points with probability approaching unity ($L_0$ consistency). It quantifies the conditions on the interplay among the averaged minimum magnitude of structural changes, the number of change points and the number of observations for consistently discovering the change points. The performance of this approach is illustrated via an analysis of a large panel of U.S. economic and financial time series data over the past 50 years.

## 1 Introduction

Data that is high dimensional and time-varying is an increasingly common occurrence in multiple domains, including economics and finance, energy generation, medical informatics, traffic engineering, audio and image processing and many others. Modeling time-series data has had a long history in statistics and econometrics, and popular methods include the autoregressive (AR) models, the integrated (I) models, and the moving average (MA) models. Most of these developments however have focused on the classical setting where the data has a fixed small dimension. A common approach in the analysis of higher-dimensional time-series data has thus been to first reduce the dimensionality (using techniques such as dynamic PCA) and then estimate time-series models over the reduced data; see the related dynamic factor model approaches in [11], [31], [32], [12], [13], [27] and [29]. Another approach is to use a static underlying model over the entire set of high dimensional time points.

In recent years, there has been a tremendous surge of research activity on the statistical estimation of (static) high-dimensional models, even under settings where the number of observations $n$ is much smaller than the number of variables $p$. It is now well understood that consistent estimation is possible even under such extreme high-dimensional settings, provided there is some underlying low-dimensional structure that can be imposed on the model space, such as sparsity, low-rank, graphical model structure and so on [34], [40] and [23]. In this paper, we focus on the high-dimensional time-series setting where the data points are high-dimensional and not necessarily independent, and the corresponding models vary with time. Across the multiple application domains mentioned above, there are observable changes over time in the overall behavior of the time-series data, indicating changes in the underlying model. Such changes in the time-series data may be gradual or abrupt depending on the application domain. In this paper, our focus is on abrupt changes, the locations of which are termed *change points*. Such a constraint of limited change points arises naturally in applications in multiple settings, including intrusion detection in computer networks, detecting downturns in economic data, detecting worsening of condition in the case of ICU patients, among other examples. Under these scenarios, using a universal (static) model to model the whole data set (without data segmentation) is not reasonable. Change point detection makes us to be able to model data and estimate parameters for each homogenous interval separately. The *goal* in this paper is to show that, given the structural constraint of magnitude and numbers of the *change points*, consistent recovery of the change points in high dimensional time series is indeed possible.

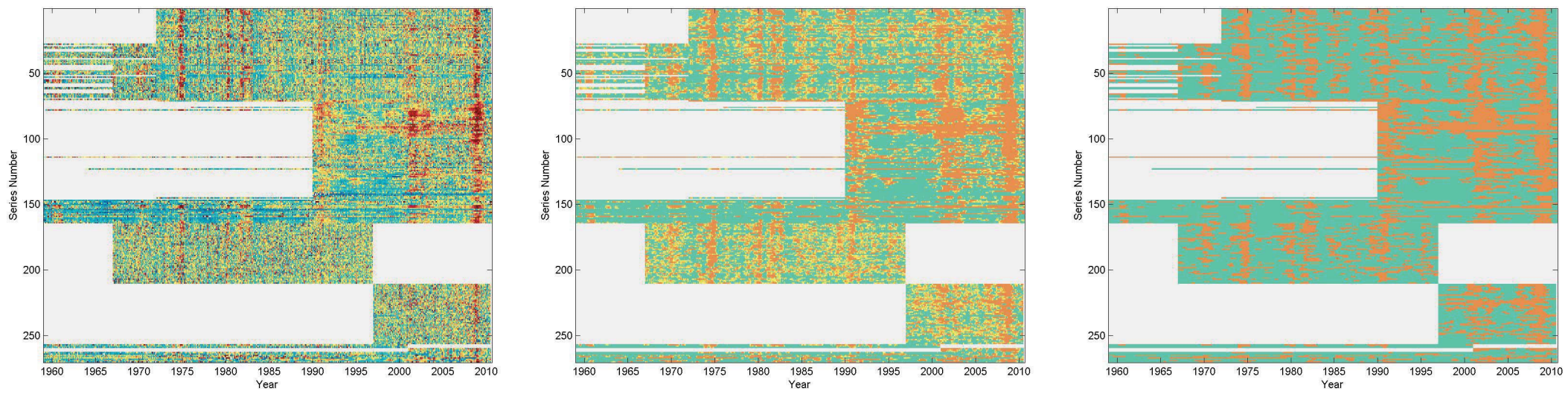

Figure 1: Heat map (with its image-enhanced version and Bry-Boschan recessions) of monthly growth rates divided by the series standard deviation for 270 series in the monthly data set as produced by [33].

A key application is the analysis of macro economic time series, that are typically measured and observed at a very low temporal frequency. Modeling and detecting structural changes is relevant for such time series since the underlying economic mechanism is likely to be disturbed by various factors. At the individual level, economic agents may change their production, consumption, savings and other patterns of behavior. This will lead to structural changes, because econometric models consist of the optimal decision rules of economic agents. A second driving force are "shocks" induced by institutional changes, such as changes of government policies from a faith in monetarist policies to reliance on more Keynesian policies. Thirdly, technological progress, which embraces the process of innovations and the long term determinants of capital accumulation, has dramatically changed economic fundamentals and economic structures in the past decades. The prevalence of structural instability in macroeconomic time series relations has indeed been confirmed by numerous empirical studies. See for instance [33], where they study the tuning points of large economic and financial data set based on some nonparametric method. Figure 1 shows the heat map (with its image-enhanced version and Bry-Boschan recessions) of monthly growth rates divided by the series standard deviation for 270 series in the monthly data set by [33]. The data set used there has 270 series; and a significant number of missing values. Blue denotes periods of positive growth, yellow denotes moderately negative growth, and red denotes strongly negative growth. The rectangular gray swaths represent missing data. The most relevant features of Figure 1 for the current purpose are the highly *synchronized* yellow-red bands. Because the horizontal axis is calendar time, the vertical yellow-red bands show periods in which many of the component series were experiencing negative growth *simultaneously*. There are two main approaches that have been popular in the detection of economic and financial change points and dating business cycles. The first is to focus on one, or atmost a few, highly aggregated time series [14] and detect change points in these. The second is to consider a large number of disparate disaggregated series, identify turning points in each of them, and then determined reference cycle change points based on the distribution of the turning points, as in [17]. In this paper, we adopt the second approach, except that we try to determine the systematic change points directly from the high dimensional time series, in the setting where the change points among the time series are synchronized.

The simplest notion of change point uses the response values themselves: in the univariate case, the time-series response is thus piece-wise constant, while in the higher-dimensional case, the different time series are not only piecewise constant, but the change points are shared across series as illustrated in the macroeconomic example above. Stipulating such a piecewise constant behavior might be too stringent for many applications: it would be much more natural to assume that the response at any time step is a linear combination of basis functions, and it is the basis coefficients that are piecewise constant, with change points shared across series. Note that this generalization would include piecewise polynomial curves, and in particular piecewise cubic splines, for instance. In this paper, we consider this general setting, where the time-series responses are linear combination of basis functions, entailing a linear regression model at any time step; across time-steps the coefficients are piecewise constant, while across series the change points are shared.

The main contribution of this paper is an analysis of the consistency in recovering the shared change points in the high-dimensional time-series data ($L_0$ consistency), as well as its $L_2$ consistency. As mentioned earlier, there has been considerable recent work in high-dimensional statistical estimation, and we now discuss connections between this previous body of work and contributions in our paper. We note that our discussion focuses mainly on work that studies high-dimensional time-varying models with change point structure. See [39], [20] for literature relating to the case of smoothly varying parameters. The authors of [20] also empirically study change point estimation in a series of Ising models, but do not present an analytical framework for it. The literature of change point detection is mainly based on the univariate iid or time series data, for example [8], [6], [5] and [21] focus on sequences of independent observations; and [1], [3], [4], [18] and others on the dependent data. These, however, provide classical asymptotics results, and do not provide finite-sample probability bounds that are typical in recent high-dimensional statistical machine learning literature, and which allow more fine-grained sample-complexity guarantees. Along the latter vein, [28] studies the recovery of piecewise constant sequence via a fused-lasso regularization as introduced in [35]: this corresponds to univariate time-series models, with iid Gaussian samples, and no basis functions. [37], an extension of the work of [15] and [16], studies a multivariate extension of this problem, specifically a piecewise constant sequence of vectors: this corresponds to multivariate time-series models, with iid Gaussian samples, and no basis functions. In [37], the change points are shared across coordinates, and hence the authors utilize a group fused Lasso regularization for the problem. They present insightful analysis for the special case of a *single* change point, but they note that the analysis of multiple change points requires a set of sophisticated tools, which they leave as an open problem. Note that both [37] and [28] do not consider the "regression" case, where the response at any coordinate and time-step is a linear combination of basis functions. The analysis of such non-identity and more generally non-orthogonal design matrix requires a subtler analysis in the high-dimensional statistical regime, and is thus considerably harder than the orthogonal case. For instance, when studying Lasso ([34]), a key part of the analysis is dealing with the non-orthogonality of the regression design matrix; and corresponding model conditions such as Restricted Isometry Property ([7]), Irrepresentability ([38]) essentially stipulating that the design matrix be suitably well-behaved. We note that regression-based change point analysis, where the signal vector is assumed to be piecewise constant, has been studied in the signal processing community, where it goes by the name of total-variation regularization and analysis regularization ([36]). This standard regression case however corresponds to univariate non-time-series and iid data. Higher-dimensional analogs of such total-variation regularization where the signal has matrix form have also been studied ([25]).

In this paper, we consider the high-dimensional time-series case where (a) we have a design matrix arising from basis functions at each time step, (b) there are multiple change points, and (c) we have dependent observations across time-steps instead of iid data as in classical statistical model estimation. As the related work discussed above indicates, each of these characteristics individually requires extending the state of the art in the analysis of high-dimensional statistical estimation. We consider an $M$-estimator that uses the group fused Lasso regularization of [37] in our regression-based setting. The contributions of this project are summarized in the following perspectives. *First*, we quantify the interplay among the number of change points, number of observations and the averaged (over different dimensions) minimum magnitude of structural changes. Specifically, the required averaged minimum magnitude of the changes should increase with fewer observations (lower $n$) and denser change points, which is intuitive from an information-theoretic perspective. *Second*, we extend the work from an iid to a time series setting by considering the general $\beta$-mixing and $\tau$-mixing time series, which includes the typical strong mixing and ARCH($\infty$) as several special cases. *Third*, as an extension, we go beyond a piecewise constant approximation of time series to a linear combination of more general basis functions, e.g. polynomial, sin and cos etc. This enables us to incorporate cases where the time series itself contains periodic variations. Our results are based on an adaptive fused group lasso framework, where the weights are adaptive over different locations. We mainly employ techniques presented in [40], [19] and [24].

The rest of the article is organized as follows. In the next section, we setup the problem with introducing main formulae, notations of temporal dependence and noise distribution assumptions. The main results concerning the $L_2$ consistency and consistency of successful change points recovery are presented in Section 3. The performance of this approach is illustrated in Section 4 via an analysis of a large panel of U.S. economic and financial time series data over the past 50 years. All technical proofs are sketched in the appendix.

# 2 Problem Setup

## 2.1 Problem Specification

We assume we have a $p$-dimensional time-series signal, with $n$ observations, $\{Y_{jt}\}_{j=1,\ldots,p,t=1,\ldots,n}$, generated as

$$Y_{jt} = \sum_{r=1}^{m} \bar{X}_{rt}\bar{\beta}_{jtr} + \varepsilon_{jt}, \tag{1}$$

where $\{\varepsilon_{jt}; j = 1,\ldots,p, t = 1,\ldots,n\}$ are noise terms that could be dependent (unlike in the iid data case), and $\{\bar{X}_{rt}; r = 1,\ldots,m; t = 1,\ldots,n\}$ are the set of features characterizing the responses $\{Y_{jt}\}$, and $\{\bar{\beta}_{jtr}\}$ are the corresponding parameters. We note that we assume that the features are shared across the coordinates $j \in \{1,\ldots,p\}$ since this is useful for most practical application. In particular, the features arise via basis functions characterizing the time-series responses, so that $\bar{X}_{rt} = f_r(t)$, for some prespecified basis functions $\{f_r(t)\}_{r=1}^m$. Typical examples of such basis functions include $f_1(t) = 1/C_1, f_2(t) = t/C_2, f_3(t) = (3t^2-1)/C_3, f_4(t) = \sin(2\pi t)/C_4$, $f_5(t) = \cos(2\pi t)/C_5, \ldots$ ($C_i$ are generic constants such that $\sum_{t=1}^n f_r^2(t)/(C_r^2 n) = 1$).

**Remark I:** It will be useful to compare the model in (1) to some of the related work on change-point detection outlined in the introduction. [28] considers the case with $p = 1, m = 1$ and $\bar{X}_{1t} = 1/n$, so that they have $Y_t = \bar{\beta}_t/n + \varepsilon_t, \quad t = 1,\ldots,n$, where $\varepsilon_t$ are assumed to be iid Gaussian. Thus, they approximate each of the high dimensional nonstationary time series $\{Y_t\}_{t=1}^m$ with a piecewise constant curves $\{\beta_t\}_{t=1}^n$. [37] considers the multivariate extension of this with $Y_{jt} = \bar{\beta}_{jt}/n + \varepsilon_{jt}, \quad t = 1,\ldots,n, j = 1,\ldots,p$, where $\varepsilon_{jt}$ are assumed to be iid Gaussian. Here, they again approximate each of the high dimensional nonstationary time series $\{Y_{jt}\}_{t=1}^m$ for $j \in \{1,\ldots,p\}$ with piecewise constant curves $\{\beta_{jt}\}_{t=1}^n$, such that the change-points are shared across the coordinates $j \in \{1,\ldots,p\}$. We note that non-time-series models with change points ([36], [25]) have the form $Y_j = \sum_{r=1}^m \bar{X}_{rj}\bar{\beta}_j + \epsilon_j$, where the signal vector $\bar{\beta}=(\bar{\beta}_1,\ldots,\bar{\beta}_m)$ is assumed to be piecewise constant along its coordinates, and the errors $\epsilon_j$ are iid Gaussian. We note that the formulation in (1) has more general characteristics than all of these: (a) it has a design matrix arising from basis functions at each time step, (b) multiple changepoints, and (c) dependent observations across time-steps instead of iid data.

### 2.1.1 A Compact Representation

Suppose we collate the responses in the $j$-th time-series: $Y_j = (Y_{j1}, Y_{j2}, \ldots, Y_{jn})$. From (1), each of these responses are given as $Y_{jt} = \sum_{r=1}^m \bar{X}_{rt}\bar{\beta}_{jtr} + \varepsilon_{jt}$. Denote

$$\bar{\bar{X}} = \begin{pmatrix} \bar{X}_{\cdot 1} & 0 & \ldots & 0 & 0 \\ 0 & \bar{X}_{\cdot 2} & \ldots & 0 & 0 \\ & \ldots & \ldots & \ldots & 0 \\ 0 & 0 & 0 & 0 & \bar{X}_{\cdot n} \end{pmatrix} \in \mathbb{R}^{n \times nm},$$

where $\bar{X}_{\cdot 1}$ is a row-vector with entries $\bar{X}_{r1}$, for $r = 1,\ldots,m$.

Suppose we collate the parameters $\bar{\beta}_j = (\bar{\beta}_{j1\cdot},\ldots,\bar{\beta}_{jn\cdot})^\top \in \mathbb{R}^{nm}$, as well as the errors $\varepsilon_j = (\varepsilon_{j1},\ldots,\varepsilon_{jn})^\top \in \mathbb{R}^n$. We can then rewrite (1) for the responses $Y_j$ in a compact matrix-theoretic notation as:

$$Y_j = \bar{\bar{X}}\bar{\beta}_j + \varepsilon_j, \tag{2}$$

Now, suppose we collate all responses $\{Y_{jt}\}$ in the vector $Y \in \mathbb{R}^{pn}$. Then, denoting $I_{p\times p}$ as the $p \times p$ identity matrix, we can further rewrite (1) for the set of all responses $Y$ as

$$Y = (I_{p\times p} \otimes \bar{\bar{X}})\bar{\beta} + \varepsilon, \tag{3}$$

with the Kronecker product $\otimes$, parameter vector $\bar{\beta}^\top = (\bar{\beta}_1^\top, \bar{\beta}_2^\top, \ldots, \bar{\beta}_p^\top)^\top$ and the error vector $\varepsilon^\top = (\varepsilon_1^\top, \varepsilon_2^\top, \ldots, \varepsilon_p^\top)^\top$.

For every *vector* $\bar{\beta} \in \mathbb{R}^{pnm}$, we also define

- $\|\bar{\beta}\|_{2,1} \stackrel{\text{def}}{=} \sum_{t=1}^n \sum_{r=1}^m \sqrt{\sum_{j=1}^p (\bar{\beta}_{jtr})^2}$ as the mixed $2-1$ norm,
- $D \stackrel{\text{def}}{=} I_{p\times p} \otimes \begin{pmatrix} 1 & 0 & \ldots & 0 & 0 \\ -1 & 1 & \ldots & 0 & 0 \\ & \ldots & \ldots & \ldots & 0 \\ 0 & 0 & 0 & -1 & 1 \end{pmatrix} \otimes I_{m\times m} \in \mathbb{R}^{pnm\times pnm}$ as the special difference matrix with $D^{-1} = I_{p\times p} \otimes \begin{pmatrix} 1 & 0 & \ldots & 0 & 0 \\ 1 & 1 & \ldots & 0 & 0 \\ & \ldots & \ldots & \ldots & 0 \\ 1 & 1 & 1 & 1 & 1 \end{pmatrix} \otimes I_{m\times m}$.

We assume the limited change points are exactly synchronized across all $p$ time series, so that $D\bar{\beta}$ is groupwise sparse. Throughout this paper, unless otherwise specified, $\|a\|$ and $\|a\|_2$ will denote the Euclidean norm of a vector $a$ (for all possible dimensions of $a$). As the grouped fused Lasso method illustrated by [37], we can estimate $\bar{\beta}$ via:

$$\begin{aligned}\widehat{\bar{\beta}} &= \underset{\bar{\beta}\in\mathbb{R}^{pnm}}{\arg\min} \frac{1}{np}\sum_{t=1}^n\sum_{j=1}^p (Y_{tj} - \sum_{r=1}^m \bar{X}_{rt}\bar{\beta}_{jtr})^2 + 2\lambda \sum_{t=1}^n\sum_{r=1}^m \sqrt{\sum_{j=1}^p (\bar{\beta}_{jtr} - \bar{\beta}_{j(t-1)r})^2} \\ &= \underset{\bar{\beta}\in\mathbb{R}^{pnm}}{\arg\min} \frac{1}{np}\|Y - (I_{p\times p}\otimes \bar{\bar{X}})D^{-1}D\bar{\beta}\|_2^2 + 2\lambda\|D\bar{\beta}\|_{2,1}. \end{aligned} \tag{4}$$

If we define $X \stackrel{\text{def}}{=} (I_{p\times p}\otimes\bar{\bar{X}})D^{-1}$ and $\beta \stackrel{\text{def}}{=} D\bar{\beta}$, then estimating $\bar{\beta}$ in (4) is equivalent to estimating $\beta$ as below:

$$\hat{\beta} = \underset{\beta\in\mathbb{R}^{pnm}}{\arg\min} \frac{1}{np}\|Y - X\beta\|_2^2 + 2\lambda\|\beta\|_{2,1}. \tag{5}$$

For a vector $\beta \in \mathbb{R}^{pnm}$, we define $\beta_{\cdot tr} = (\beta_{1tr}, \beta_{2tr}, \ldots, \beta_{ptr})^\top, 1 \leqslant t \leqslant n, 1 \leqslant r \leqslant m$, which we often rewrite as $\beta_i, 1 \leqslant i \leqslant nm$ for simplicity of notation. If $\hat{\beta}_i$ represents the estimate, and $\beta_i^*$ the oracle parameter, then we define

$$\widehat{\mathcal{M}} = \mathcal{M}(\hat{\beta}) \stackrel{\text{def}}{=} \{1 \leqslant i \leqslant nm : \|\hat{\beta}_i\|_2 \neq 0\}, \qquad \mathcal{M}^* = \mathcal{M}(\beta^*) \stackrel{\text{def}}{=} \{1 \leqslant i \leqslant nm : \|\beta_i^*\|_2 \neq 0\}.$$

Thus $\mathcal{M}(\hat{\beta})$ is the set of change points recovered by the estimate $\hat{\beta}$ and $\mathcal{M}(\beta^*)$ is the oracle set of change points. Finally, $\mathcal{M}^c$ denotes the complement of the set of indices $\mathcal{M}$.

### 2.2 Distributional Assumptions on the Noise Coefficients

In this subsection, we set out to outline two measures of temporal dependence, namely $\beta$-mixing and $\tau$-mixing for the consistency analysis w.r.t. $m \geqslant 1$ and $m = 1$ respectively. For the latter (and yet special) case $m = 1$, due to the special structure of the $X$ matrix, we are able to use some special inequality to obtain a sharper $L_2$ bound. For our first result for the general case $m \geqslant 1$ in Theorem 1, we need the following definition of $\beta$-mixing. Suppose we have a sequence of random variables $\{A_t\}_{t=-\infty}^{\infty}$ where each $A_t$ is a measurable function from a probability space $(\Omega, \mathcal{F}, \mathrm{P})$ into a measurable space $\mathcal{A}$. Following the notation of [22], we use $\mathbf{A_i^j}$ to denote a block of this random sequence $\{A_t\}_{t=i}^{j}$ where $i$ and $j$ are integers that may be infinite, and use similar notation for the $\sigma$-fields generated by these blocks and their joint distributions. In particular, $\sigma_i^j$ denotes the $\sigma$-field generated by $\mathbf{A_i^j}$, and $\mathrm{P}_i^j$ denotes the joint distribution of $\mathbf{A_i^j}$. Following [10], we state the definitions related to $\beta$-mixing below:

**Definition 2.1** ($\beta$-mixing Coefficient)**.** *For each positive integer t, the $\beta$-mixing coefficient $\beta(t)$ is*

$$\beta(t) = \sup_a \|\, \mathrm{P}_{-\infty}^a\, \mathrm{P}_{a+t}^{\infty} - \mathrm{P}_{a,t}\, \|_{TV}, \tag{6}$$

*where $\|\cdot\|_{TV}$ is the total variation norm, $\otimes$ is the Kronecker product, and $\mathrm{P}_{a,t}$ is the joint distribution of $(\mathbf{A}_{-\infty}^{\mathbf{a}}, \mathbf{A}_{\mathbf{a+t}}^{\infty})$. A stochastic process is said to be $\beta$-mixing if $\beta(t) \to 0$ as $t \to \infty$.*

Here, depending on context, we use $\beta(t)$ to denote the $\beta$-mixing coefficient. This should not be confused with the parameter $\beta$ in (5). Loosely stated, the $\beta$-mixing coefficient $\beta(t)$ measures the total-variation distance between the joint distribution of random variables separated by $t$ time units and the distribution under which random variables separated by $t$ time units are independent.

Now, consider the model in (5). We assume that for any $j$, the $\beta$-mixing sequence $\{\varepsilon_{jt'}\}_{t'=1}^{n}$ satisfies the following assumptions.

A1 $\forall t',\ \mathsf{E}\,\varepsilon_{jt'} = 0$.

A2 $\exists \sigma^2, \forall t_1', t_2', (t_2')^{-1}\,\mathsf{E}(\varepsilon_{jt_1'} + \ldots + \varepsilon_{j(t_1'+t_2')})^2 \leqslant \sigma^2$.

A3 $\forall t', |\varepsilon_{jt'}| \leqslant M$.

A4 For any $\varepsilon > 0$, and $\theta \overset{\text{def}}{=} \varepsilon^2/4$, the $\beta$-mixing coefficient satisfies

$$\beta([q\theta n/(1+\theta)] - 1) = \mathcal{O}\{((npm)^{2+\delta'}\sqrt{\log(npm)n})^{-1}\},$$

with $\delta' > 0$, $q\theta T/(1+\theta) = 3\theta/(1+\theta) = 3\theta\sigma^2/(Ms_n)$ and square brackets denoting the integer part.

Assumptions [A1-A3] come from [10][P.35], and in general terms, these require $X_{jtm}\varepsilon_{jt}$ to be mean zero, have a finite second moment and bounded. This is similar to the bounded second moment requirement in the iid scenario. Assumption [A2] relates to stationarity. Assumption [A4] qualifies the degree of the temporal dependence in the form of the $\beta$-mixing coefficient.

In practice, not all time series are $\beta$-mixing, and [9] introduces the notion of $\tau$-mixing, which covers several typical time series frequently observed in economics and finance (discussed later in Remark II). Thus, for our second result for the special case $m = 1$ in Theorem 2, we state the definitions related to $\tau$-mixing below.

For any real random variable $X$ in $L^1$ and any $\sigma$-algebra $M$ of $\mathcal{F}$, let $\mathrm{P}_{X|M}$ be a conditional distribution distribution of $X$ given $M$ and let $\mathrm{P}_X$ be the distribution of $X$. The coupling coefficient $\tau(M, X)$ is a measure of weak dependence ([9]) given by

$$\tau(M, X) = \Big\| \sup_{f \in \wedge_1(\mathbb{R})} \Big| \int f(x)\,\mathrm{P}_{X|M}(dx) - \int f(x)\,\mathrm{P}_X(dx) \Big| \Big\|_1,$$

where $\wedge_1(\mathbb{R})$ is the set of 1-Lipschitz functions from $\mathbb{R}$ to $\mathbb{R}$. The multivariate analogue is defined as follows: If $Y$ is a random variable with values in $\mathbb{R}^k$, the coupling coefficient $\tau$ is defined as follows: $\tau(M, Y) = \sup\{\tau(M, f(Y)), f \in \wedge_1(\mathbb{R}^k)\}$, where $\wedge_1(\mathbb{R}^k)$ is the set of 1-Lipschitz functions from $\mathbb{R}^k$ to $\mathbb{R}$.

**Definition 2.2** ($\tau$-mixing Coefficient)**.** *The $\tau$-mixing coefficients $\tau(t)$ of a sequence $\{A_t\}_{t=1}^{n}$ of real-valued random variables are defined by $\tau(t) = \sup_{k \geqslant 0} \tau_k(t)$ where $\tau_k(t) = \max_{1 \leqslant l \leqslant k} \frac{1}{l} \sup_{\hat{p},(j_1,\ldots,j_l)}\{\tau(\sigma_1^{\hat{p}}, (A_{j_1}, \ldots, A_{j_l})), \hat{p} + t \leqslant j_1 < \ldots < j_l\}$.*

Intuitively, like the $\beta$-mixing coefficient $\beta(t)$, the $\tau$-mixing coefficient $\tau(t)$ also measures the distance between random variables separated by $t$ time units. We overload notation to use $\tau(t) = \tau([t])$ (square brackets denoting the integer part). Now, consider the model in (5) for $m = 1$. We assume that for any $j$, the $\tau$-mixing sequences $\{A_t\}_{t=1}^{n} = \{\varepsilon_{jt}\}_{t=1}^{n}$ satisfies the following assumptions.

A5 There exists a positive constant $\gamma_1$ such that the $\tau$-mixing coefficient satisfies

$$\tau(\varepsilon) \leqslant \exp(-c\varepsilon^{\gamma_1}) \text{ for any } \varepsilon \geqslant 1, \text{ where } c > 0 \tag{7}$$

A6 There exists a positive constant $\gamma_2$ such that

$$\sup_{t>0} \mathrm{P}(|\varepsilon_t| > x) \leqslant \exp(1 - x^{\gamma_2}) \overset{\text{def}}{=} H(x) \text{ for any positive } x \tag{8}$$

A7 Suppose $\gamma_1$ and $\gamma_2$ are the constants outlined in Assumptions [A5] and [A6]. Then,

$$\gamma < 1 \text{ where } \frac{1}{\gamma} = \frac{1}{\gamma_1} + \frac{1}{\gamma_2}. \tag{9}$$

It is also useful to define an auto-correlation-like term $V$ to understand the results presented in the next section, as below:

$$V = \sup_{M \geqslant 1} \sup_{t>0} \{\mathsf{Var}(\varphi_M(\varepsilon_t)) + 2 \sum_{j>t} |\,\mathsf{Cov}(\varphi_M(\varepsilon_t), \varphi_M(\varepsilon_j))|\}. \tag{10}$$

where for any positive $M$, $\varphi_M(\varepsilon) = (\varepsilon \wedge M) \vee (-M)$.

**Remark II:** The assumptions here require that the distribution of $\varepsilon_t$ *not* to be fat-tailed. As discussed in [24], the set of $\tau$-mixing time series satisfying (7), (8) and (9) cover a wide variety of time-series. Examples include $\alpha$(strong)-mixing time series (provided the $\alpha$ mixing coefficients satisfy certain conditions), instantaneous functions of absolutely regular processes, functions of linear processes with absolutely regular innovations, as well as error sequences of ARCH($\infty$) models.

## 3 Main Results and Their Implications

In this section, we present statistical guarantees associated with our estimator, where we provide results on its $L_2$ consistency, as well as its sparsistency ($L_0$ consistency). A key quantity is the number of change points $s \stackrel{\text{def}}{=} \mathcal{M}(\beta^*)$. We also denote $\phi_{max}$ as the maximum eigenvalue of the matrix $X^\top X/n$. And we use the restricted strong convexity (RSC) condition ([26]) defined as

$$\kappa(s) := \min \left\{ \frac{\|X\Delta\|_2}{\sqrt{n} \parallel \Delta \parallel_2} : \Delta \in \mathbb{R}^{nmp} \backslash \{0\}, \parallel \Delta_{\mathcal{M}_c(\beta^*)} \parallel_{2,1} \leqslant 3 \parallel \Delta_{\mathcal{M}(\beta^*)} \parallel_{2,1} \right\}, \tag{11}$$

where $\Delta_{\mathcal{M}(\beta^*)}$ denotes the vector formed by stacking the rows of matrix $\Delta$ w.r.t. the index set $\mathcal{M}(\beta^*)$.

### 3.1 $L_2$-Consistency

**Theorem 1** ($L_2$-Consistency for $m \geqslant 1$)**.** *Let $\{\varepsilon_{jt'}\}_{t'=1}^n$ (for each $j$) be a sequence of centered real valued $\beta$-mixing random variables satisfying assumptions* $[A1] - [A4]$*. Suppose we set the regularization penalty as $\lambda = M'\sqrt{\log(npm)/np}$. Then, for sufficiently large $M'$ depending on $\varepsilon, \sigma^2, M$ (as in Assumption A3), with probability at least* $1 - (npm)^{-M'^2} - (npm)^{-(2+\delta')}$*, any solution $\hat{\beta}$ of* (5) *satisfies:*

$$\frac{1}{np}\|X(\hat{\beta} - \beta^*)\|_2^2 \leqslant \frac{16M'^2 s \log{(npm)}}{\kappa^2 n} \tag{12}$$

$$\frac{1}{\sqrt{p}}\|\hat{\beta} - \beta^*\|_{2,1} \leqslant \frac{16M' s\sqrt{\log{(npm)}}}{\kappa^2\sqrt{n}} \tag{13}$$

$$\mathcal{M}(\hat{\beta}) \leqslant \frac{64\phi_{\max}}{\kappa^2} s \tag{14}$$

For the next theorem, we first define $u(n) \stackrel{\text{def}}{=} \max\{(M_1 \log n)^{1/\gamma}, \{M_2(1 + n\,V)n^{\delta}\}^{1/2}, (M_3\, n^{1+\delta})^{1/2}\}$ for some constant $\delta > 0$ and constants $M_1, M_2, M_3$.

**Theorem 2** ($L_2$-Consistency for $m = 1$)**.** *Let $\{\varepsilon_{jt}\}_{t=1}^n$ (for each $j$) be a sequence of centered real valued $\tau$-mixing random variables with $V$ defined as in* (10)*. Assume that Assumptions [A5]-[A7] hold. Suppose we set the regularization penalty as $\lambda = \frac{2u(n)}{\sqrt{pn}}$. Then, $V$ is finite, and with probability* *at least* $1-(np)^{1-M_1}-p^{1-M_2}-p^{1-M_3}$, *any solution* $\hat{\beta}$ *of* (5) *satisfies:*

$$\frac{1}{np}\|X(\hat{\beta}-\beta^*)\|_2^2 \leqslant \frac{64su^2(n)}{\kappa^2n^2} \tag{15}$$

$$\frac{1}{\sqrt{p}}\|\hat{\beta}-\beta^*\|_{2,1} \leqslant \frac{32su(n)}{\kappa^2n} \tag{16}$$

$$\mathcal{M}(\hat{\beta}) \leqslant \frac{64\phi_{\max}}{\kappa^2}s \tag{17}$$

**Remark III:** The difference between the special case $m=1$ and $m \geqslant 1$ is an artifact of the technical tools employed in our analysis. However, since $u(n)$ is of the order of $\sqrt{n}$, the right-hand-side (RHS) of (12) and (13) are of the order of $\log n/n$ and $\sqrt{\log n/n}$ respectively and the RHS of (15) and (16) possess tighter bounds, with order behavior $1/n$ and $\sqrt{1/n}$ respectively.

### 3.2 Consistent Recovery of Change Points

For the change point analysis, the most important target is the successful recovery of the change point set. Thus, in this section, we develop sufficient conditions for the consistency of change point recovery. To this end, we consider the case that the tuning parameter $\lambda$ may not be universal, but could change with different locations, i.e. we introduce additional weights $w_{tr}$'s for different $\beta_{tr}$'s ($tr=1,\dots,nm$). Note that, if we choose all $w_{tr}=1$, this corresponds to the stand group fused Lasso regularization as in the last section. Similar to (5), we redefine $\hat{\beta}$ (not be confused with that in the last section) as

$$\hat{\beta} = \underset{\beta\in\mathbb{R}^{pnm}}{\arg\min}\frac{1}{np}\|Y-X\beta\|^2 + 2\lambda\sum_{t=1}^{n}\sum_{r=1}^{m}w_{tr}\sqrt{\sum_{j=1}^{p}(\bar{\beta}_{jtr})^2}. \tag{18}$$

Note that, if all $w_{tr}=1$ (the previous case), all derivations are still valid as before. We implement a two-step procedure. As mentioned before, for notational convenience, we rewrite $w_{tr},\beta_{tr}, 1\leqslant tr\leqslant nm$ as $w_i,\beta_i, 1\leqslant i\leqslant nm$. For simplicity, we focus on the case when $m=1$, noting that an extension to $m\geqslant 1$ is fairly straightforward. We define $b_{\mathcal{M}^*}=\min\{\|\beta_i^*\|_2/\sqrt{p} : i\in\mathcal{M}^*\}$ and thus $b_{\mathcal{M}^*}$ can be interpreted as the averaged (over different dimensions) minimum magnitude of the changes. Let the matrix $X_{\mathcal{M}^*}$ be that submatrix whose columns are indexed by $\mathcal{M}^*$, the vector $\beta^*_{\mathcal{M}^*}$ is the reduced dimensional vector built upon the components of $\beta^*$ indexed by $\mathcal{M}^*$, $\Psi_{\mathcal{M}^*,\mathcal{M}^*} \overset{\text{def}}{=} \frac{X_{\mathcal{M}^*\top}X_{\mathcal{M}^*}}{n}$, $\phi_1$ be the smallest eigenvalue of $\Psi_{\mathcal{M}^*,\mathcal{M}^*}$ and $\phi_2$ is the maximum eigenvalue $\frac{X_{\mathcal{M}_c^*\top}}{\sqrt{n}}\frac{X_{\mathcal{M}^*}}{\sqrt{n}}(\Psi_{\mathcal{M}^*,\mathcal{M}^*})^{-1}X_{\mathcal{M}^*\top}$. Note that this involves the covariance between the submatrix $X_{\mathcal{M}^*}$ w.r.t. the oracle set and the submatrix $X_{\mathcal{M}_c^*\top}$ w.r.t. the empty set. Also define $S_{\mathcal{M}^*}=(\frac{w_i\beta_i^*}{\|\beta_i^*\|}, i\in\mathcal{M}(\beta^*))$. We use the $sgn(\cdot)$ function which maps positive entry to 1, negative entry to $-1$ and zero to 0 and $(\cdot)=_s(\cdot)$ abbreviates $sgn(\cdot)=sgn(\cdot)$.

**Theorem 3.** *Suppose that the following assumptions are satisfied with a high probability:*

*B1* $b_{\mathcal{M}^*}\geqslant\frac{2u(n)}{n\phi_1}$, *B2* $\lambda\leqslant(\frac{b_{\mathcal{M}^*}\phi_1}{2\sqrt{p w_i}})$, *B3* $\lambda\geqslant\frac{2u(n)\phi_2}{n\sqrt{pw_i(1-k)}}, \forall i\in\mathcal{M}_c^*$,

*B4* $\|\{\frac{(X^\top)_i}{\sqrt{n}}\frac{X_{\mathcal{M}^*}}{\sqrt{n}}(\Psi_{\mathcal{M}^*,\mathcal{M}^*})^{-1}\tilde{S}_{\mathcal{M}^*}\}_i\|_2 < kw_i, \forall i\in\mathcal{M}_c^*$.

*We have* $\forall 1\leqslant i\leqslant nm$,

$$\mathrm{P}(\|\hat{\beta}_i\|_2 =_s \|\beta_i^*\|_2)=1.$$

**Remark IV:** Assumption [B1] is a constraint on the averaged (over different dimensions) minimum magnitude of the changes, which is a decreasing function of $n$ and $\phi_1$ and a increasing function of $u(n)$. Since a smaller $\phi_1$ indicates that the change points are closer together and larger $u(n)$ indicates there are more change points, to correctly recover them, the minimum magnitude of the changes themselves must be larger. Assumptions [B2-B3] impose requirements on the tuning parameter $\lambda$. Assumption [B4] states that "low correlation between variables in $\mathcal{M}_c^*$ and variables in $\mathcal{M}^*$" is necessary for the group fused Lasso method to select the change points consistently and this can be understood as the "strong irrepresentable condition" from [38] and [2].

| Year | Month |
|---|---|
| 1974 | 08 |
| 1977 | 01 |
| 1979 | 09 |
| 1981 | 08 |
| 1982 | 11 |
| 1984 | 10 |
| 1986 | 12 |
| 1990 | 02 |
| 2003 | 12 |

| Year | Month |
|---|---|
| 1964 | 12 |
| 1970 | 11 |
| 1973 | 06 |
| 1974 | 11 |
| 1977 | 01 |
| 1979 | 09 |
| 1981 | 10 |
| 1983 | 01 |
| 1986 | 04 |
| 1990 | 09 |
| 1997 | 01 |
| 2001 | 05 |
| 2003 | 12 |

| Year | Month |
|---|---|
| 1964 | 12 |
| 1970 | 11 |
| 1973 | 06 |
| 1974 | 11 |
| 1977 | 01 |
| 1979 | 09 |
| 1981 | 08 |
| 1983 | 01 |
| 1985 | 11 |
| 1990 | 11 |
| 1997 | 01 |
| 2001 | 05 |
| 2003 | 12 |

| Year | Month |
|---|---|
| 1964 | 12 |
| 1970 | 11 |
| 1973 | 06 |
| 1974 | 11 |
| 1977 | 01 |
| 1979 | 09 |
| 1981 | 08 |
| 1983 | 01 |
| 1985 | 11 |
| 1990 | 11 |
| 1997 | 01 |
| 2001 | 05 |
| 2003 | 12 |

Table 1: Change points discovered using this method with $\alpha = 0, 1/2, 1, 2$ (from left to right).

Additionally,

$$\mathrm{P}\left( \|\{ \frac{(X^\top)_i}{\sqrt{n}} \frac{X_{\mathcal{M}^*}}{\sqrt{n}} (\Psi_{\mathcal{M}^*,\mathcal{M}^*})^{-1} \tilde{S}_{\mathcal{M}^*} \}_i \|_2 \leqslant w_i, \forall i \in \mathcal{M}_c^* \right) \tag{19}$$

is a necessary condition (*weaker* than the sufficient condition) for the change point selection. This can be shown by contradiction. If $\|\{ \frac{(X^\top)_i}{\sqrt{n}} \frac{X_{\mathcal{M}^*}}{\sqrt{n}} (\Psi_{\mathcal{M}^*,\mathcal{M}^*})^{-1} \lambda \tilde{S}_{\mathcal{M}^*} \}_i \|_2 > \lambda w_i, \forall i \in \mathcal{M}_c^*$, and if $\frac{1}{np} \|\{ X^\top (I_{np \times np} - X_{\mathcal{M}^*} (\Psi_{\mathcal{M}^*,\mathcal{M}^*})^{-1} \frac{X_{\mathcal{M}^*\top}}{n}) \varepsilon \}_i \|_2 < \|\{ \frac{(X^\top)_i}{\sqrt{n}} \frac{X_{\mathcal{M}^*}}{\sqrt{n}} (\Psi_{\mathcal{M}^*,\mathcal{M}^*})^{-1} \lambda \tilde{S}_{\mathcal{M}^*} \}_i \|_2 - \lambda w_i, \forall i \in \mathcal{M}_c^*$ holds with a high probability, which is guaranteed by the concentration inequality, then we have $\forall i \in \mathcal{M}_c^*$, $\frac{1}{np} \|\{ X^\top (Y - X\hat{\beta}) \}_i \| > \lambda w_i$, which contradicts the KKT condition.

In summary, returning to (19), we see that if we were to increase the weight $w_i$ for $i \in \mathcal{M}_c^*$ and decrease the weights otherwise, we can ensure consistency. This, however, requires the (potentially empirical) knowledge of $\mathcal{M}^*$, which is exactly the idea underlying an adaptive scheme for the problem.

## 4 Empirical Study

We use the large panel dataset of economic and financial time series of US over the past 50 years for illustrative purposes. This dataset from [30] contains 131 monthly macro indicators covering a broad range of categories including income, industrial production, capacity, employment and unemployment, consumer prices, producer prices, wages, housing starts, inventories and orders, stock prices, interest rates for different maturities, exchange rates, money aggregates and so on., over a January 1960 to December 2003 time-span. We apply the group LARS algorithm based on [37] with the weights $w_n = 1/\beta_{OLS}^\alpha$, where $\beta_{OLS}$ is the OLS estimate for $\beta$, and $\alpha$ is varied from 0 (w.r.t. the unweighted penalty) to $1/2, 1, 2$ (w.r.t. the data dependant penalty). We showcase our results in Table 1. Comparing the change points obtained here with the yellow-red bands in Figure 1 as cataloged by empirical economists, we find a large match between the two if we use the data dependent penalty, while the unweighted one leads to a large difference from the empirical observations as in Figure 1.

**Proof of Theorem 1:** We divide the proof into two parts: non-probability part and probability part.

**Non-probability part:** If $\frac{2}{pn}\|X^\top\varepsilon\|_{2,\infty} \leqslant \lambda$ holds, then we could obtain the following intermediate results from a more unified approach as in [26].

$$\frac{1}{pn}\|X(\hat{\beta}-\beta^*)\|^2 \leqslant \frac{16sp}{\kappa^2}\lambda^2; \tag{20}$$

$$\frac{1}{\sqrt{p}}\sum_{i=1}^{nm}\|\hat{\beta}_i-\beta_i^*\|_2 \leqslant \frac{16s\sqrt{p}}{\kappa^2}\lambda. \tag{21}$$

For this particular problem, following their notations, if we choose $S_{\mathcal{G}} = \mathcal{M}(\beta^*)$ (thus $s_{\mathcal{G}} = s$), the loss function $\mathcal{L} = \frac{1}{np}\|Y - X\beta\|^2$, $\mathcal{R} = 2\lambda\|\beta\|_{2,1}$ and $n = np$ here, Corollary 4 of [26] show that, for this particular example, if $\frac{2}{pn}\|X^\top\varepsilon\|_{2,\infty} \leqslant \lambda$ and assumption (11) holds, then we have (24) and (25).

For $\beta^* \in \mathbb{R}^{npm}$, by the definition of $\hat{\beta}$ in (5) and $Y = X\beta^* + \varepsilon$, we have

$$\frac{1}{pn}\|X\hat{\beta}-Y\|^2 + 2\lambda\sum_{i=1}^{nm}\|\hat{\beta}_i\|_2 \leqslant \frac{1}{pT}\|X\beta^*-Y\|^2 + 2\lambda\sum_{i=1}^{nm}\|\beta_i^*\|_2$$

$$\frac{1}{pn}\|X(\hat{\beta}-\beta^*)\|^2 \leqslant \frac{2}{pn}\underbrace{\varepsilon^\top X(\hat{\beta}-\beta^*)}_{\leqslant\|X^\top\varepsilon\|_{2,\infty}\|\hat{\beta}-\beta^*\|_{2,1}} + 2\lambda\sum_{i=1}^{nm}(\|\beta_i^*\|_2 - \|\hat{\beta}_i\|_2)$$

If $\frac{2}{pn}\|X^\top\varepsilon\|_{2,\infty} \leqslant \lambda$ holds with a high probability, then we have

$$\frac{1}{pn}\|X(\hat{\beta}-\beta^*)\|^2 \leqslant \lambda\sum_{i=1}^{nm}\|\hat{\beta}_i-\beta_i^*\|_2 + 2\lambda\sum_{i=1}^{nm}(\|\beta_i^*\|_2 - \|\hat{\beta}_i\|_2)$$

$$\frac{1}{pn}\|X(\hat{\beta}-\beta^*)\|^2 + \lambda\sum_{i=1}^{nm}\|\hat{\beta}_i-\beta_i^*\|_2 \leqslant 4\lambda\sum_{i\in\mathcal{M}(\beta^*)}\|\hat{\beta}_i-\beta_i^*\|_2 \tag{22}$$

Then we know

$$\lambda\sum_{i=1}^{nm}\|\hat{\beta}_i-\beta_i^*\|_2 \leqslant 4\lambda\sum_{i\in\mathcal{M}(\beta^*)}\|\hat{\beta}_i-\beta_i^*\|_2; \sum_{i\in\mathcal{M}(\beta^*)^c}\|\hat{\beta}_i-\beta_i^*\| \leqslant 3\sum_{i\in\mathcal{M}(\beta^*)}\|\hat{\beta}_i-\beta_i^*\|$$

Thus, by the assumption (11), we have

$$\|(\hat{\beta}-\beta^*)^{\mathcal{M}(\beta^*)}\|_2 \leqslant \frac{\|X(\hat{\beta}-\beta^*)\|}{\kappa\sqrt{n}}. \tag{23}$$

By coming (22) and (23), we have

$$\frac{1}{pn}\|X(\hat{\beta}-\beta^*)\|^2 \leqslant 4\lambda\sum_{i\in\mathcal{M}(\beta^*)}\|\hat{\beta}_i-\beta_i^*\|_2 \leqslant 4\lambda\sqrt{s}\|(\hat{\beta}-\beta^*)^{\mathcal{M}(\beta^*)}\|_2 \leqslant 4\lambda\sqrt{s}\frac{\|X(\hat{\beta}-\beta^*)\|_2}{\kappa\sqrt{n}}.$$

Solving this inequality w.r.t. $\|X(\hat{\beta}-\beta^*)\|$ gives $\|X(\hat{\beta}-\beta^*)\| \leqslant \frac{4\lambda\sqrt{s}p\sqrt{n}}{\kappa}$ and thus

$$\frac{1}{pn}\|X(\hat{\beta}-\beta^*)\|^2 \leqslant \frac{16sp}{\kappa^2}\lambda^2; \tag{24}$$

$$\begin{aligned}\frac{1}{\sqrt{p}}\sum_{i=1}^{nm}\|\hat{\beta}_i-\beta_i^*\|_2 &\leqslant \frac{4}{\sqrt{p}}\sum_{i\in\mathcal{M}(\beta^*)}\|\hat{\beta}_i-\beta_i^*\|_2 \leqslant \frac{4\sqrt{s}\|(\hat{\beta}-\beta^*)^{\mathcal{M}(\beta^*)}\|_2}{\sqrt{p}}\\ &\leqslant \frac{4\sqrt{s}\|X(\hat{\beta}-\beta^*)\|}{\kappa\sqrt{np}} = \frac{16s\sqrt{p}}{\kappa^2}\lambda. \end{aligned}\tag{25}$$

If $\hat\beta_i \neq 0$ and $\frac{2}{pn}\|X^\top \varepsilon\|_{2,\infty} \leqslant \lambda$ holds with a high probability, we have

$$\frac{1}{np}\{X^\top(Y - X\hat\beta)\}_i = \lambda\frac{\hat\beta_i}{\|\hat\beta_i\|}$$
$$\|\frac{1}{np}\{X^\top X(\beta^* - \hat\beta) + X^\top\varepsilon\}_i\|_2 = \lambda$$
$$\|\frac{1}{np}\{X^\top X(\beta^* - \hat\beta)_i\}\|_2 \geqslant \lambda - \|\frac{1}{Tp}(X^\top\varepsilon)_i\|_2 \geqslant \lambda/2$$
$$\sum_{i\in\mathcal{M}(\hat\beta)}\|\frac{1}{np}\{X^\top X(\beta^* - \hat\beta)_i\}\|_2^2 \geqslant \frac{\lambda^2}{4}\mathcal{M}(\hat\beta)$$

and thus

$$\begin{aligned}\mathcal{M}(\hat\beta) &\leqslant \frac{4}{\lambda^2p^2n^2}\sum_{i=1}^{nm}\|\{X^\top X(\beta^* - \hat\beta)_i\}\|^2\\ &\leqslant \frac{4\phi_{max}}{\lambda^2np^2}\|X(\hat\beta - \beta^*)\|^2 \leqslant \frac{4\phi_{max}}{\lambda^2Tp^2}\frac{16\lambda^2sp^2T}{\kappa^2} \leqslant \frac{64\phi_{max}s}{\kappa^2}\end{aligned} \tag{26}$$

**Probability part:** In the following, we study the probability $\frac{2}{pn}\|X^\top\varepsilon\|_{2,\infty} > \lambda$. Since $X_{j't',jtr} = 0$ for $j' \neq j$, we have

$$\mathrm{P}\Big\{\max_{1\leqslant tr\leqslant nm}\sqrt{\sum_{j=1}^{p}(\sum_{t'=1}^{n}\sum_{j'=1}^{p}X_{j't',jtr}\varepsilon_{j't'})^2} > \frac{\lambda pn}{2}\Big\} = \mathrm{P}\Big\{\max_{1\leqslant tr\leqslant nm}\sqrt{\sum_{j=1}^{p}(\sum_{t'=1}^{n}X_{jt',jtr}\varepsilon_{jt'})^2} > \frac{\lambda pn}{2}\Big\}$$
$$\leqslant \mathrm{P}\Big\{\sqrt{\sum_{j=1}^{p}\max_{1\leqslant tr\leqslant nm}(\sum_{t'=1}^{n}X_{jt',jtr}\varepsilon_{jt'})^2} > \frac{\lambda pn}{2}\Big\} = \mathrm{P}\Big\{\max_{1\leqslant j\leqslant p}\max_{1\leqslant tr\leqslant nm}(\sum_{t'=1}^{n}X_{jt',jtr}\varepsilon_{jt'})^2 > \frac{\lambda^2pn^2}{4}\Big\}$$
$$\leqslant p\,\mathrm{P}\Big\{\max_{1\leqslant tr\leqslant nm}|\sum_{t'=1}^{n}X_{jt',jtr}\varepsilon_{jt'}| > \frac{\lambda\sqrt{p}n}{2}\}$$

**For $m = 1$:** Given the special structure of $X = X^oD^{-1} = (I_{p\times p} \otimes I_{n\times n})(I_{p\times p} \otimes \begin{pmatrix}1 & 0 & \dots & 0 & 0\\ 1 & 1 & \dots & 0 & 0\\ & \dots & \dots & \dots & 0\\ 1 & 1 & 1 & 1 & 1\end{pmatrix})$, if $m = 1$, it is not hard to see that $\max_{1\leqslant tr\leqslant nm}|\sum_{t'=1}^{n}X_{jt',jtr}\varepsilon_{jt'}| = \max_{1\leqslant t\leqslant n}|\sum_{t'=t}^{n}\varepsilon_{jt'}|, \forall j$. Thus, by [24]

$$p\,\mathrm{P}\Big(\max_{1\leqslant t\leqslant n}|\sum_{t'=t}^{n}\varepsilon_{jt'}| > \frac{\lambda\sqrt{p}n}{2}\Big) = p\,\mathrm{P}\Big(\max_{1\leqslant t\leqslant n}|\sum_{t'=1}^{t}\varepsilon_{jt'}| > \frac{\lambda\sqrt{p}n}{2}\Big)$$
$$\leqslant np\exp\Big(-\frac{u^\gamma}{C_1}\Big) + p\exp\Big\{-\frac{u^2}{C_2(1+nV)}\Big\} + p\exp\Big[-\frac{u^2}{C_3n}\exp\Big\{\frac{u^{\gamma(1+\gamma)}}{C_4(\log u)^\gamma}\Big\}\Big],\ \textit{with}\quad u \overset{\text{def}}{=} \frac{\lambda\sqrt{p}n}{2}$$

If we choose $\frac{\lambda\sqrt{p}n}{2} = u(n) = \max\{(M_1\log np)^{1/\gamma}, \{M_2(1+nV)\log p\}^{1/2}, (M_3n\log p^{1/2}\}$ where $M_1, M_2, M_3$ are some large enough constants depending on $C_1, C_2, C_3$ respectively and $\delta > 0$, then $\|X^\top\varepsilon\|_{2,\infty} > \frac{pn\lambda}{2}$ holds with a probability at least $1 - (np)^{1-M_1} - p^{1-M_2} - p^{1-M_3}$. (15) and (16) follow by plugging $\lambda$ into (24) and (25).

For $m \geqslant 1$:,

$$\mathrm{P}\Big\{\max_{1\leqslant tr\leqslant nm}\sqrt{\sum_{j=1}^{p}(\sum_{t'=1}^{n}X_{jt',jtr}\varepsilon_{jt'})^2} > \frac{\lambda pn}{2}\Big\}$$

$$\leqslant nm\,\mathrm{P}\Big\{\sum_{j=1}^{p}(\sum_{t'=1}^{n}X_{jt',jtr}\varepsilon_{jt'})^2 > \frac{\lambda^2p^2n^2}{4}\Big\} \leqslant nm\,\mathrm{P}\Big\{\max_{1\leqslant j\leqslant p}(\sum_{t'=1}^{n}X_{jt',jtr}\varepsilon_{jt'})^2 > \frac{\lambda^2pn^2}{4}\Big\}$$

$$\leqslant npm\,\mathrm{P}\Big\{|\sum_{t'=1}^{n}\varepsilon_{jt'}| > \frac{\lambda\sqrt{pn}}{2}\Big\}.$$

Assume that the $\beta$-mixing sequence $\{\varepsilon_{jt'}\}_{t'=1}^n$ satisfying assumption [A1-A3] for all $\forall j$, applying the Bernstein type inequality for $\beta$-mixing random variables $\{\varepsilon_{jt'}\}_{t'=1}^n$ (see [10], Theorem 4 on [P.36]) yields that, $\forall\varepsilon > 0$ ($\theta \overset{\mathrm{def}}{=} \varepsilon^2/4$) and $\forall\ 0 < q \leqslant 1$ and $o \overset{\mathrm{def}}{=} \frac{\lambda\sqrt{p}}{2}$,

$$\mathrm{P}(|\sum_{t'=1}^{n}\varepsilon_{jt'}| \geqslant on) \leqslant \underbrace{4\exp\Big[-\frac{(1-\varepsilon)3(1+\theta)o^2n}{2\{3(1+\theta)\sigma^2+qMon\}}\Big]}_{\overset{\mathrm{def}}{=}A} + \underbrace{2\,\frac{(1+\theta)\beta([q\theta n/(1+\theta)]-1)}{q}}_{\overset{\mathrm{def}}{=}B}.$$

To make $npm(A+B)$ arbitrarily small, we choose $\frac{\lambda\sqrt{p}}{2} = o = M'\sqrt{\log(npm)/n}$ with sufficiently large $M'(\varepsilon,\sigma^2,M)$, which also depends on $\varepsilon,\sigma^2,M$, $\log(npm)/n = \mathcal{O}(1)$, $q = 3(1+\theta)\sigma^2/(Mon)$, and $\beta([q\theta n/(1+\theta)]-1) = \mathcal{O}\{((npm)^{2+\delta'}\sqrt{\log(nm)n})^{-1}\}$ with $\delta' > 0$ and $q\theta n/(1+\theta) = 3\theta\sigma^2/(Mo)$. Thus parts $A$ and $B$ are bounded by $\exp(-M'^2\log(npm))$ and $(npm)^{-(2+\delta')}$ respectively, which can be arbitrarily close to 0. This completes the proof. □

**Proof of Theorem 3:** For $\hat\beta$ defined in (18), KKT conditions give:

$$\text{If } \|\hat\beta_i\| \neq 0,\ \frac{1}{np}\{X^\top(Y-X\hat\beta)\}_i = \lambda\frac{w_i\hat\beta_i}{\|\hat\beta_i\|} \quad (27)$$

$$\text{If } \|\hat\beta_i\| = 0,\ \frac{1}{np}\|\{X^\top(Y-X\hat\beta)\}_i\|_2 \leqslant \lambda w_i \quad (28)$$

And

$$\begin{aligned}\hat\beta_{\mathcal{M}^*} &= \Big(\frac{X_{\mathcal{M}^*}^\top X_{\mathcal{M}^*}}{np}\Big)^{-1}\Big(\frac{X_{\mathcal{M}^*}^\top Y}{np} - \lambda S_{\mathcal{M}^*}\Big)\\ &= \beta^*_{\mathcal{M}^*} + \Big(\frac{X_{\mathcal{M}^*}^\top X_{\mathcal{M}^*}}{np}\Big)^{-1}\Big(\frac{X_{\mathcal{M}^*}^\top \varepsilon}{np} - \lambda S_{\mathcal{M}^*}\Big)\\ &= \beta^*_{\mathcal{M}^*} + (\Psi_{\mathcal{M}^*,\mathcal{M}^*})^{-1}\Big(\frac{X_{\mathcal{M}^*}^\top \varepsilon}{n} - p\lambda S_{\mathcal{M}^*}\Big).\end{aligned}$$

If $\hat\beta_{\mathcal{M}^*} =_s \beta^*_{\mathcal{M}^*}$, then (27) and (28) hold for $\hat\beta = (\hat\beta^\top_{\mathcal{M}^*}, 0^\top_{\mathcal{M}^*_c})$. Since $X\hat\beta = X_{\mathcal{M}^*}\hat\beta_{\mathcal{M}^*}$ for this $\hat\beta$, we have, $\forall 1 \leqslant i \leqslant nm$,

$$\|\hat\beta_i\| =_s \|\beta^*\| \text{ if } \begin{cases} \|\hat\beta_i\| =_s \|\beta^*_i\|, \forall i \in \mathcal{M}^* \\ \frac{1}{np}\|\{X^\top(Y-X\hat\beta)\}_i\|_2 \leqslant \lambda w_i, \forall i \in \mathcal{M}^*_c \end{cases}$$

Since $Y - X\hat\beta = \varepsilon - X_{\mathcal{M}^*}(\hat\beta_{\mathcal{M}^*} - \beta^*_{\mathcal{M}^*}) = \varepsilon - X_{\mathcal{M}^*}(\Psi_{\mathcal{M}^*,\mathcal{M}^*})^{-1}(\frac{X^\top_{\mathcal{M}^*}\varepsilon}{n} - p\lambda S_{\mathcal{M}^*}), \forall 1 \leqslant i \leqslant nm$,

$$\|\hat\beta_i\| =_s \|\beta^*\| \text{ if } \begin{cases} \|\hat\beta_i\| =_s \|\beta^*_i\|, \forall i \in \mathcal{M}^* \\ \frac{1}{np}\|\{X^\top(I - X_{\mathcal{M}^*}(\Psi_{\mathcal{M}^*,\mathcal{M}^*})^{-1}\frac{X^\top_{\mathcal{M}^*}}{n})\varepsilon + X^\top X_{\mathcal{M}^*}(\Psi_{\mathcal{M}^*,\mathcal{M}^*})^{-1}p\lambda S_{\mathcal{M}^*}\}_i\| \leqslant \lambda\ \forall i \in \mathcal{M}^*_c. \end{cases}$$

Then we know that

$$
\begin{aligned}
&\mathrm{P}(\|\hat{\beta}_i\| \neq_s \|\beta_i^*\|, \forall 1 \leqslant i \leqslant nm) \\
\leqslant &\mathrm{P}\left(\|(e_i \otimes I_{p\times p})^\top (\Psi_{\mathcal{M}^*,\mathcal{M}^*})^{-1} \frac{X_{\mathcal{M}^*}^\top \varepsilon}{n}\|_2 \geqslant \|\beta_i^*\|_2/2, \forall i \in \mathcal{M}^*\right) && (29)\\
+ &\mathrm{P}\left(\|(e_i \otimes I_{p\times p})^\top (\Psi_{\mathcal{M}^*,\mathcal{M}^*})^{-1} p\lambda S_{\mathcal{M}^*}\|_2 \geqslant \|\beta_i^*\|_2/2, \forall i \in \mathcal{M}^*\right) && (30)\\
+ &\mathrm{P}\left(\frac{1}{np}\|\{X^\top (I_{np\times np} - X_{\mathcal{M}^*}(\Psi_{\mathcal{M}^*,\mathcal{M}^*})^{-1} \frac{X_{\mathcal{M}^*}^\top}{n})\varepsilon\}_i\|_2 \geqslant (1-k)\lambda w_i, \forall i \in \mathcal{M}_c^*\right) && (31)\\
+ &\mathrm{P}\left(\frac{1}{np}\|\{X^\top X_{\mathcal{M}^*}(\Psi_{\mathcal{M}^*,\mathcal{M}^*})^{-1} p\lambda S_{\mathcal{M}^*}\}_i\|_2 \geqslant \lambda k w_i, \forall i \in \mathcal{M}_c^*\right) && (32)
\end{aligned}
$$

for any $0 < k < 1$ and $e_i$ is the unit vector of length $|\mathcal{M}^*|$ with the $m$th entry nonzero. We now need bound each of (29) - (32).

**For** (29), from the definitions of $b_{\mathcal{M}^*}$ and $\phi_1$, we have

$$
\begin{aligned}
&\mathrm{P}\left(\|(e_i \otimes I_{p\times p})^\top (\Psi_{\mathcal{M}^*,\mathcal{M}^*})^{-1} \frac{X_{\mathcal{M}^*}^\top \varepsilon}{n}\|_2 \geqslant \|\beta_i^*\|_2/2, \forall i \in \mathcal{M}^*\right) \\
\leqslant &\mathrm{P}\left(\|X_i^\top \varepsilon\|_2 \geqslant b_{\mathcal{M}^*} n\sqrt{p}\phi_1/2, \forall i \in \mathcal{M}^*\right),
\end{aligned}
$$

which we have studied before. If we make $b_{\mathcal{M}^*} n\sqrt{p}\phi_1/2 \geqslant \frac{pn\lambda}{2} = \sqrt{p}u(n)$, then we could get the requirement on the averaged (over different dimensions) minimum magnitude of the changes as the assumption [B1]: $b_{\mathcal{M}^*} \geqslant \frac{2u(n)}{n\phi_1}$, which is, not surprisingly, a decreasing function of $n$.

**For** (30), we have

$$
\begin{aligned}
&\mathrm{P}\left(\|(e_i \otimes I_{p\times p})^\top (\Psi_{\mathcal{M}^*,\mathcal{M}^*})^{-1} p\lambda S_{\mathcal{M}^*}\|_2 \geqslant \|\beta_i^*\|_2/2, \forall i \in \mathcal{M}^*\right) \\
\leqslant &\mathrm{P}\left(\lambda \geqslant \frac{b_{\mathcal{M}^*}\phi_1}{2\sqrt{p}w_i}, \forall i \in \mathcal{M}^*\right),
\end{aligned}
$$

which means $\lambda$ need satisfy $\lambda \leqslant \mathcal{O}_p(\frac{b_{\mathcal{M}^*}\phi_1}{2\sqrt{p}w_i})$ with a high probability.

**For** (31), $\mathrm{P}\left(\frac{1}{np}\|(X^\top \varepsilon)_i\|_2 \geqslant \frac{(1-k)\lambda w_i}{2}, i \in \mathcal{M}_c^*\right)$ has been studied before.

$$
\begin{aligned}
&\mathrm{P}\left(\frac{1}{np}\|\{X^\top (X_{\mathcal{M}^*}(\Psi_{\mathcal{M}^*,\mathcal{M}^*})^{-1} \frac{X_{\mathcal{M}^*}^\top}{n})\varepsilon\}_i\|_2 \geqslant (1-k)\lambda w_i/2, \forall i \in \mathcal{M}_c^*\right) \\
= &\mathrm{P}\left(\frac{1}{np}\|\frac{X_i^\top}{\sqrt{n}}\frac{X_{\mathcal{M}^*}}{\sqrt{n}}(\Psi_{\mathcal{M}^*,\mathcal{M}^*})^{-1} X_{\mathcal{M}^*}^\top \varepsilon\|_2 \geqslant (1-k)\lambda w_i/2, \forall i \in \mathcal{M}_c^*\right) \\
\leqslant &\mathrm{P}\left(\frac{1}{np}\|(X_i^\top \varepsilon)\|_2 \geqslant \frac{(1-k)\lambda w_i}{2\phi_2}, \forall i \in \mathcal{M}_c^*\right).
\end{aligned}
$$

If we make $\frac{(1-k)np\lambda w_i}{2\phi_2} = \sqrt{p}u(n)$ and thus $\lambda \geqslant \frac{2u(n)\phi_2}{n\sqrt{pw_i(1-k)}}, \forall i \in \mathcal{M}_c^*$.

**For** (32), $\forall 0 < k < 1$, we have

$$
\begin{aligned}
&\mathrm{P}\left(\frac{1}{np}\|\{X^\top X_{\mathcal{M}^*}(\Psi_{\mathcal{M}^*,\mathcal{M}^*})^{-1} p\lambda S_{\mathcal{M}^*}\}_i\|_2 \geqslant k\lambda w_i, \forall i \in \mathcal{M}_c^*\right) \\
= &\mathrm{P}\left(\|\{\frac{X_i^\top}{\sqrt{n}}\frac{X_{\mathcal{M}^*}}{\sqrt{n}}(\Psi_{\mathcal{M}^*,\mathcal{M}^*})^{-1} S_{\mathcal{M}^*}\}_i\|_2 \geqslant k w_i, \forall i \in \mathcal{M}_c^*\right),
\end{aligned}
$$

which is guarantied by the assumption [B4]. □